\documentclass[5p]{elsarticle}
\usepackage{setspace}
\usepackage{lineno}
\modulolinenumbers[5]
\usepackage{amssymb}
\usepackage{amsmath}
\usepackage[colorlinks=true,linkcolor=blue,urlcolor=blue,citecolor=blue]{hyperref}
\biboptions{numbers,sort&compress}
\journal{Journal of Magnetism and Magnetic Materials}
\begin{document}

\begin{frontmatter}

\title{Melting of Ferromagnetic Order on a Trellis Ladder}

\author{Debasmita Maiti } 

\author{Manoranjan Kumar\corref{mycorrespondingauthor}}
\cortext[mycorrespondingauthor]{Corresponding author}
\ead{manoranjan.kumar@bose.res.in}
\address{S. N. Bose National Centre for Basic Sciences, Block JD, Sector III, Salt Lake, Kolkata - 700106, India}

\date{\today}
\begin{abstract}
The ground state properties of a frustrated spin-1/2 system is studied on a trellis ladder which is composed 
of two zigzag ladders interacting through rung interactions. The presence of rung interaction between the
zigzag ladders induces a non-magnetic ground state, although, each of zigzag ladders has ferromagnetic order in 
 weak anti-ferromagnetic leg interaction limit. The rung interaction also generates rung dimers and 
opens spin gap which increases rapidly with rung interaction strength. 
The correlation between spins decreases exponentially with the distance between them.     

\end{abstract}


\end{frontmatter}


\section{\label{sec:intro}Introduction}

The interaction-driven quantum phase transition in low-dimensional 
frustrated systems like spin chain ~\cite{chubukov1991,furukawa2012}, ladder or any quasi-one dimensional system has been
a frontier area of current research ~\cite{white1994,verkholyak2012}. Many realizations of 
these systems like (N$_2$H$_5$)CuCl$_3$~\cite{Maeshima2003},
LiCuSbO$_4$~\cite{dutton_prl}, LiCuVO$_4$~\cite{Mourigal2012}, Li$_2$CuZrO$_4$~\cite{Drechsler2007} 
are frustrated due to competing spin exchange interactions. $J_1-J_2$ Heisenberg spin-1/2 model 
with nearest neighbor (NN) ferromagnetic (FM) $J_1$ and next nearest neighbor (NNN) anti-ferromagnetic (AFM) 
$J_2$ interactions is used extensively to understand the ground state magnetic 
properties of many of these materials~\cite{white96,OKAMOTO1992,Chitra1995,mk2015}. This model can explain the gapless spin fluid,
 gapped dimer and incommensurate gapped spiral phases ~\cite{white96,mk2010a,mk2015}. On the 
other hand, the ground state of quasi-1D ladder like structure with AFM leg and 
rung interactions shows the presence of gapped short-range order in the system. 
This kind of phase is observed in SrCu$_2$O$_3$~\cite{sandvik1995}, 
(VO)$_2$P$_2$O$_7$ ~\cite{dcjhon1987,dagotto96}, 
CaV$_2$O$_5$, MgV$_2$O$_5$ ~\cite{Korotin1999,Tdasgupta2000} etc.

\begin{figure}
\centering
\includegraphics[width=3.0 in]{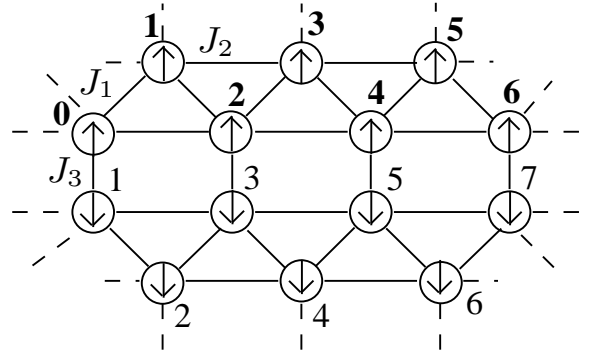}
\caption{\label{fig:figure1} Two coupled zigzag ladders form trellis
ladder. The broken lines show the extension of trellis ladder to trellis lattice structure. The arrows show respective spin arrangements.
The reference site is labeled by '0' and the distances of all other sites are 
shown with respect to it. The distance of the sites in the same zigzag ladder 
as the reference site are written in bold numbers, and the sites on the other ladder are written in normal numbers. }
\end{figure}

The $J_1-J_2$ chain can also be considered as a two-chain lattice with diagonal or zigzag like
couplings. Hence it can be alternatively called as zigzag ladder. The isolated ladders like zigzag ~\cite{Chitra1995,white96} and 
normal ladder ~\cite{white96,white1994} have been extensively studied. However, the theoretical study of the effect of
inter-ladder coupling on the ladders is still an open field. One of the extended networks 
of the coupled ladders can form a trellis lattice like structure. The trellis lattice is composed of a number of 
normal ladders coupled through zigzag bonds; alternatively, it can be considered as coupled zigzag 
ladders through rung interactions as shown in Fig.~\ref{fig:figure1}. In this lattice, 
$J_2$ and $J_3$ are leg and rung couplings of a normal ladder, respectively and $J_1$ is inter-ladder coupling through zigzag bonds. 
   
There are several theoretical studies for two coupled zigzag ladders, e.g. a two-leg honeycomb ladder is considered in Ref.~\cite{Luo2018}, 
where both $J_1$ and $J_2$ are AFM, but $J_3$ can be either FM or 
AFM. This system shows two types of Haldane phases for the FM $J_3$ 
and, columnar dimer and rung singlet phases for the AFM $J_3$.
Normand \textit{et al.} considered the similar coupled ladders with all three AFM $J_1$, $J_2$ and $J_3$ interactions.
They find dimerized chains for large $J_2$ 
and small $J_3$ limit, spiral long range order for both large $J_2$ and $J_3$ limit, N\'eel long range order in 
the small $J_2 < 0.4$ and for all $J_3$ ~\cite{Normand1997}. Ronald \textit{et al.} have shown the effect of 
inter-chain coupling on spiral ground state of $J_1-J_2$ model ~\cite{Zinke2009}. 
The effect of inter-ladder coupling on spin gap and magnon dispersion 
has been discussed in ~\cite{Miyahara1998} exploiting a theoretical model which has also been compared with the experimental data 
of SrCu$_2$O$_3$ and CaV$_2$O$_5$. The study of interladder coupling effect is important to explain the physical properties of 
some other materials like LaCuO$_{2.5}$~\cite{Troyer1997}, Sr$_{14}$Cu$_{24}$O$_{41}$~\cite{Uehara1996}, MgV$_2$O$_5$~\cite{Korotin1999}, 
NaV$_2$O$_5$~\cite{smolinski1998,Tanokura2010} etc. The weak interladder exchange interactions in these materials
form an effective 2D trellis like structure. Recently Yamaguchi \textit{et al.} showed magnetic field induced spin nematic phase in the 
verdazyl radical $\beta$-2,3,5-$Cl_3$-V~\cite{Yamaguchi2018}. The $J_1$-$J_2$-$J_3$ spin system promises a zoo of exotic phases.  

In the current paper we consider a trellis 
ladder which is composed of two zigzag ladders with FM $J_1$ and AFM $J_2$ and coupled by AFM $J_3$ 
as shown in Fig.~\ref{fig:figure1}. The ground state of an isolated zigzag ladder in small $J_2$ 
limit is FM. Our main focus of this paper is to understand the effect of rung interaction 
$J_3$ on the FM ground state exhibited by a single zigzag ladder in the limit of 
$J_2 \le \frac{|J_1|}{4}$, and study the transition of ground state from the FM to the singlet dimer state.

This paper is divided into four sections. In section~\ref{sec2}, the model Hamiltonian 
and the numerical methods are explained. The numerical results are given in section~\ref{sec3}. 
All the results are discussed and summarized in section~\ref{sec4}. 

\section{\label{sec2}Model Hamiltonian and Numerical Method}
We consider four-leg ladder system where two zigzag ladders are coupled through 
an AFM  Heisenberg interaction as shown in Fig.~\ref{fig:figure1}. The diagonal interactions 
$J_1$ in zigzag ladders are FM whereas $J_2$ bonds along the legs are AFM. 
Two zigzag ladders interact with each other through AFM rungs $J_3$. 
Thus we can write an isotropic Heisenberg spin-1/2 model Hamiltonian for the system as 

\begin{eqnarray}
\label{eq:ham}
H &= &\sum_{a = 1,2}\sum_{i=1}^{N/2} \ J_1 \, \vec{S}_{a,i} \cdot \vec{S}_{a,i+1} 
+ J_2 \, \vec{S}_{a,i} \cdot \vec{S}_{a,i+2} \nonumber \\
& & \qquad \qquad + J_3 \, \vec{S}_{1,i} \cdot \vec{S}_{2,i},
\end{eqnarray}
where $a=1,2$ are the zigzag ladder indices. $\vec{S}_{a,i}$ is the spin operator at reference site $i$ on zigzag ladder $a$. We set the FM $J_1=-1$ and treat the AFM $J_2$ and $J_3$ as variable quantities.
We use periodic boundary condition along the rungs, whereas it is 
open along the legs of the system.   

We use the exact diagonalization (ED) method for small systems and density matrix renormalization 
group (DMRG) method to handle the large degrees of freedom 
for large systems. The DMRG is a state of art numerical technique for 1D or quasi-1D system, and 
it is based on the systematic truncation of irrelevant degrees of freedom at every step of growth 
of the chain~\cite{white-prl92,karen2006,schollwock2005}. We use recently developed DMRG method where 
four new sites are added at every DMRG steps~\cite{mk2010b}. For the renormalization of operators, we keep
 $m$ eigenvectors corresponding to largest eigenvalues of the density matrix of the system in 
the ground state of the Hamiltonian in Eq.~(\ref{eq:ham}). We have kept $m$ up to 300 to constraint the truncation error 
less than $10^{-10}$. We have used system sizes up to $N=400$ to minimize 
the finite size effect.  

\section{\label{sec3}Results}
It is well known that the zigzag spin-1/2 ladder has a FM ordered ground state for FM $J_1$ and AFM
$J_2 \le \frac{|J_1|}{4}$ ~\cite{mk2015}. The AFM coupling, i.e., the rung interaction $J_3$ between two isolated zigzag spin-1/2 ladders
retains the FM arrangement in each zigzag ladder; however, the spins on two zigzag ladders are 
in AFM arrangement to each other, as depicted in the schematic Fig.~\ref{fig:figure1}. In other words, the effective ground state
of the whole system 
is in $S^z=0$ manifold, though the spins on each zigzag ladder are arranged ferromagnetically.
The ground state forms singlet dimer along the rung in the large $J_3$ limit, and ground state wave 
function can be represented as the product of singlet dimers. 
However, to study the effect of $J_3$ 
on the ground state, we analyze longitudinal correlation function, longitudinal 
bond order and spin gap.             

\begin{figure}
\centering
\includegraphics[width=3.5 in]{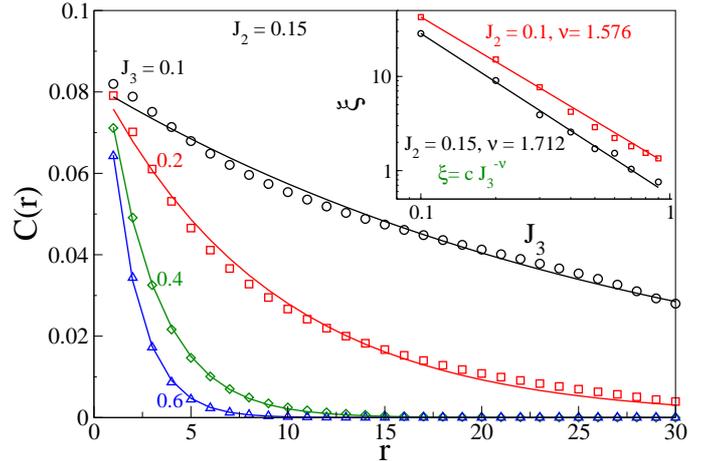}
\caption{\label{fig:cor}The longitudinal spin-spin correlation C(r) shown for N = 122, $J_2=0.15$ and different $J_3$ as 
indicated adjacent the respective curves. The solid lines represent respective exponential fits. 
The inset shows the correlation length $\xi$ vs. $J_3$ plots for $J_2=0.1$ and $0.15$ in log-log scale.
}
\end{figure}

In this paper we focus $J_2/|J_1| < \frac{1}{4}$ limit where each zigzag ladder have the FM
order in the ground state for $J_3=0$. For a finite $J_3$, we calculate the longitudinal 
spin-spin correlations $C(r)=<{S^z}_i {S^z}_{i+r}>$, where $S^z_i$ and $S_{i+r}^z$ are the z-component of the spin operators 
at the reference site $i$ and the site at a distance $r$ from $i$, respectively. 
In \ref{fig:figure1}, We have shown the distance $r$ along the same zigzag ladder with bold numerics, and otherwise in normal numerics,
where the reference site is at the 0th position. We note that in $J_2/|J_1|< \frac{1}{4}$ limit all the spins are aligned in parallel 
on each zigzag ladder and have short range longitudinal correlation for finite $J_3$. 
As we increase the strength of $J_3$, $C(r)$ shows an exponential behavior as shown in the 
main Fig.~\ref{fig:cor} for $J_2 = 0.15$. We notice that each zigzag ladder shows FM
arrangement as $C(r)>0$, 
but it decays exponentially with $r$, i.e. $ C(r) \propto exp(-r/\xi)$. 
The correlation length $\xi$ follows 
an algebraic decay with $J_3$ for a given $J_2$, as shown in the inset of Fig. \ref{fig:cor}. 
$\xi$ for $J_3=0.1$ is approximately 28.5, however it decrease to 1.58 for $J_3=0.6$. 
$\xi$ becomes less than 1 for $J_3 > 0.9$ for $J_2=0.15$, and in this parameter regime the system 
is completely dimerized along the rung.

\begin{figure}
\centering
\includegraphics[width=3.3 in]{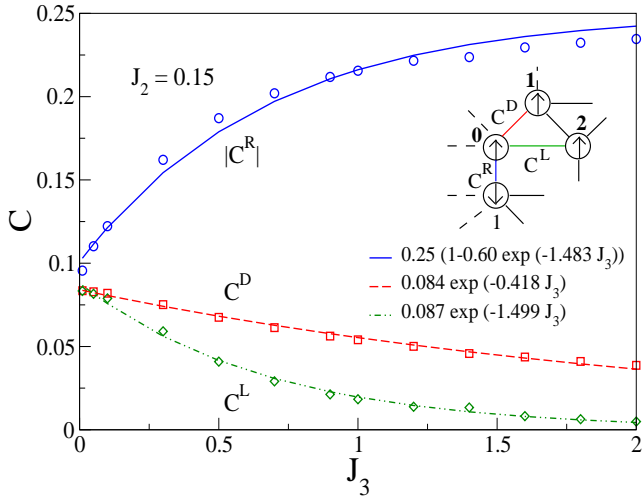}
\caption{\label{fig:cor1} 
Correlation function C of the reference site with its first neighbors along the rung ($C^R$, circle), 
diagonal direction ($C^D$, square), and the leg ($C^L$, diamond), are shown as function of $J_3$. We have drawn $|C^R|$ 
to take care of the AFM $J_3$ interaction. The lines represent respective exponential fits. 
}
\end{figure}

To study the effect of $J_3$ on the bonds along the rung, diagonal and leg directions,
we calculate correlations $C^R$, $C^D$ and $C^L$, respectively as shown in Fig.~\ref{fig:cor1}. 
The calculations of these three correlations are 
confined to the first neighbor along the respective directions.
$|C^R|$ increases exponentially with $J_3$ and follows $|C^R| = 0.25(1-0.60exp(-1.48J_3))$. 
$C^D$ and $C^L$ are represented by square and diamond symbols, respectively, and 
both these bond orders exponentially decrease with $J_3$. 
The exponent of the $C^D$ and $C^L$ are 0.42 and 1.50 respectively. $C^L$ decays faster than
$C^D$, because $J_2$ allows the magnon to deconfine along the leg of the zigzag ladder;
therefore weaker $J_2$ reduces $C^L$.

\begin{figure}
\centering
\includegraphics[width=3.5 in]{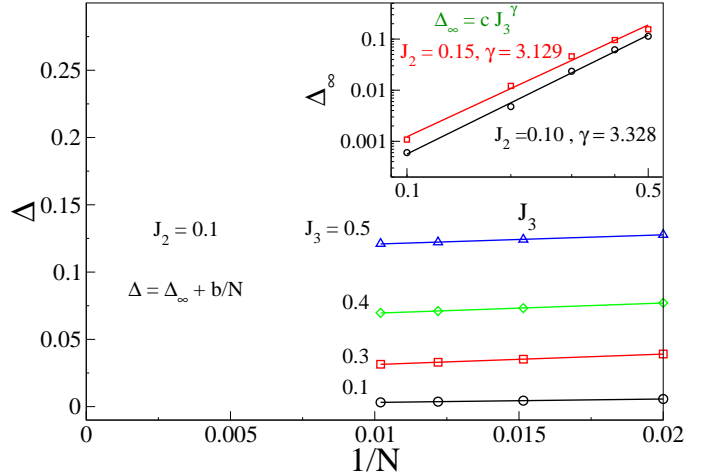}
\caption{\label{fig:spin_gap} The extrapolation of spin-gap $\Delta$ with respect to system size $N$
for different $J_3=0.1, 0.3, 0.4$ and 0.5 at $J_2=0.1$. The solid lines are the fitted curves.
In the inset, the spin gap in thermodynamic limit,
$\Delta_\infty$ vs. $J_3$ plots for $J_2=0.1$ and 0.15 are shown in log-log scale.}
\end{figure}

The correlation function $C(r)$ of the system shows 
the short range spin order. Therefore, we explore the excitation energy or spin 
gap in the system. The rung interaction dominates other interactions; thus we expect the opening of 
the spin gap $\Delta$. We calculate $\Delta$ for various $J_3=0.1, 0.3, 0.4$ and 0.5. The main 
Fig. \ref{fig:spin_gap} shows the extrapolation of the spin gap $\Delta$, from which we obtain the spin gap $\Delta_\infty$ in the
thermodynamic limit. $\Delta_\infty$ increases algebraically with $J_3$, as shown in the inset of 
Fig. \ref{fig:spin_gap} for $J_2=0.1$ and 0.15.
The algebraic exponent $\gamma$ for $J_2=0.1$ and 0.15 are 3.33 and 3.13, respectively.
We notice that $\gamma$ decreases with increasing $J_2$. 
This may be due to the delocalization of magnon along the leg of zigzag ladder.  
The large $J_2/|J_1|$ enhances the interaction of spin along each leg of zigzag ladder and 
the each leg of the system can have quasi-long range correlation like 
a normal Heisenberg chain, in $J_2/|J_1|>>1$ limit. In this limit, system 
behaves as four non-interacting spin-1/2 Heisenberg chains with gapless spectrum ~\cite{mk2016}.

\section{\label{sec4}Discussion and Conclusions}
In this paper the effect of $J_3$ on FM order in each zigzag ladder 
of a trellis ladder with FM $J_1$ and AFM $J_2 < \frac{|J_1|}{4}$ is studied. 
We show that even a small $J_3$ induces spin gap in the system. The correlation between 
spins on a zigzag ladder decays exponentially with distance. It may be because of the confinement 
of magnon along the rungs. As shown in Fig. \ref{fig:spin_gap}
the gap increases rapidly with $J_3$ for $J_2 = 0.1$. The correlation length of the
system reduced to less than a unit lattice for $J_3 > 0.9$ at $ J_2 \approx 0.15$. This implies the setting of the dimerized state.      

This model can also be mapped to a two interacting $J_1-J_2$ Heisenberg spin-1/2 chains. This system 
is studied recently by Ronald \textit{et al.}~\cite{Zinke2009}. They have mostly studied the 
effect of inter-chain coupling on the spiral nature of the ground state in large 
$J_2$ and low $J_3$ limit. There are many compounds like 
CaV$_2$O$_5$ ~\cite{ONODA1996}, SrCu$_2$O$_3$ ~\cite{Miyahara1998,dagotto96} etc, 
which have both strong $J_3$ and $J_2$. However, our prediction are confined 
to the $J_2 < \frac{|J_1|}{4}$ and large $J_3$ limit.      

In summary, this model system goes from a FM ordered ground state along a zigzag ladder in the $J_3=0$ limit 
to a rung dimer state in large $J_3$ limit. The correlation length $\xi$ of the system decreases algebraically 
and spin gap $\Delta_\infty$ increases algebraically with exponent higher than $\gamma>3$, on increase in $J_3$ for a given $J_2$. 

\textbf{Acknowledgements} MK thanks DST for a Ramanujan Fellowship SR/S2/RJN-69/2012
and DST for funding computation facility through SNB/MK/14-15/137.

\end{document}